\documentclass[10pt]{article}

\usepackage{amsmath,amssymb,amsthm,amscd,amsfonts}
\usepackage[round,authoryear]{natbib}
\usepackage[text={17cm,24cm}]{geometry}
\usepackage{graphicx}

\newcommand{\bcd}{\textit{bcd}}

\newcommand{\Drosophila}{\textit{Drosophila}}

\begin{document}

\title{Two-exponential models of gene expression patterns for noisy experimental data}
\author{Theodore Alexandrov\,$^{\text{1,2,3}}$, Nina Golyandina\,$^{\text{4,}*}$,\\ David Holloway\,$^{\text{5}}$, Alex Shlemov\,$^{\text{4}}$ and Alexander Spirov\,$^{\text{6,7}}$ }
\maketitle

\date{\noindent$^{\text{1}}$EMBL Heidelberg, Meyerhofstr. 1, Heidelberg, 69117, Germany,\\
$^{\text{2}}$Skaggs School of Pharmacy and Pharmaceutical Sciences, University of California of San Diego, La Jolla, CA 9500, USA,\\
$^{\text{\sf 3}}$SCiLS GmbH, Bremen, 28359, Germany,\\
$^{\text{\sf 4}}$St. Petersburg State University, Universitetskaya nab. 7/9, St.Petersburg, 199034, Russia,\\
$^{\text{\sf 5}}$Mathematics Department, British Columbia Institute of Technology, 3700 Willingdon Avenue, Burnaby, B.C., V5G 3H2, Canada,\\
$^{\text{\sf 6}}$Computer Science and CEWIT, SUNY Stony Brook, 1500 Stony Brook Road, Stony Brook, NY 11794, USA and\\
$^{\text{\sf 7}}$The Sechenov Institute of Evolutionary Physiology \& Biochemistry, Torez Pr. 44, St.Petersburg, 194223, Russia.\\
$*$ the corresponding author, nina@gistatgroup.com}

\abstract{\textbf{Motivation:} Spatial pattern formation of the primary anterior-posterior morphogenetic gradient of the transcription factor Bicoid (Bcd) has been studied experimentally and computationally for many years. Bcd specifies positional information for the downstream segmentation genes, affecting the fly body plan. More recently, a number of researchers have focused on the patterning dynamics of the underlying \bcd{} mRNA gradient, which is translated into Bcd protein. New, more accurate techniques for visualizing \bcd{} mRNA need to be combined with quantitative signal extraction techniques to reconstruct the \bcd{} mRNA distribution. \\
\textbf{Results:} Here, we present a robust technique for quantifying gradients with a two-exponential model. This approach: 1) has natural, biologically relevant parameters; and 2) is invariant to linear transformations of the data which can arise due to variation in experimental conditions (e.g. microscope settings, non-specific background signal). This allows us to quantify \bcd{} mRNA gradient variability from embryo to embryo (important for studying the robustness of developmental regulatory networks); sort out atypical gradients; and classify embryos to developmental stage by quantitative gradient parameters.

\section{Background}
\label{sec:intro}

\textbf{Biology} A key concept in developmental biology is that of morphogen gradients \citep{Briscoe.etal2010}, in which a spatially-distributed gradient of a signaling molecule (morphogen) affects downstream cellular responses in a concentration-dependent manner. These spatial gradients are established by molecular transport, either active or diffusional. One of the best-studied morphogen gradients in development is of the protein transcription factor Bicoid (Bcd) \citep{Briscoe.etal2010,Grimm.etal2010}, which regulates gene expression along the anterior-posterior (AP) axis of the developing fruit fly (\Drosophila{}) embryo. The Bcd protein gradient has been studied quantitatively for many years, both in terms of quantitative experiments and in mathematical modeling of the dynamics of gradient formation.

More recently, studies have focused on the underlying dynamics and patterning of the \bcd{} mRNA gradient, since the Bcd protein forms via translation from the mRNA. The \bcd{} RNA gradient forms earlier than the protein gradient, and exhibits a number of distinct features from the protein pattern. These have been the subject of several mathematical modeling projects, as well as new quantitative experimental projects to characterize the \bcd{} mRNA gradient \citep{Spirov.etal2009,Lipshitz2009,Kavousanakis.etal2010,Deng.etal2010,Little.etal2011,Cheung.etal2011,Cheung.etal2014,Dalessi2012130,
Liu.Niranjan2012,Fahmy.etal2014, Ali-Murthy.Kornberg2016}.

There are a number of features to \bcd{} RNA patterning which make it more complex to study than the Bcd protein pattern. These features require new and more sophisticated techniques in data acquisition and signal processing in order to extract quantitative data. The current paper presents and validates a new method for quantitative analysis of spatial profiles of \bcd{} RNA reliably extracted from whole embryo 3D scans (confocal microscopy) of FISH (fluorescent in situ hybridization) RNA data.

\smallskip
\textbf{Data} Figure~\ref{fig3}A shows a sagittal section through the middle of such a whole embryo scan, with fluorescence intensity proportional to the concentration of \bcd{} mRNA. The dataset is 3D, and the RNA transport setting up this 3D pattern has components in the three coordinates: head-to-tail (AP); top-to-bottom (dorso-ventral, DV); and inside-to-outside (basal-apical, BA). The gradient is chiefly along the AP direction: biologically, the mRNA spreads posteriorly from a maternal deposition at the anterior end of the embryo. There are however, concentration differences in the DV direction, and while \bcd{} RNA and protein patterns are most intense in the surface, or cortex, of the embryo, \bcd{} is also found in the interior of the embryo, and there is a concentration gradient in the BA direction. The transport processes establishing these gradients may differ between the different coordinates: AP transport of \bcd{} RNA involves minus-end motors trafficking along microtubules, assisted by proteins such as Staufen (Stau) \citep{Weil.etal2006,Weil.etal2008,Spirov.etal2009,Fahmy.etal2014, Ali-Murthy.Kornberg2016}; DV `bending' of \bcd{} pattern may reflect geometric asymmetries in the embryo; and BA transport appears to occur at later stages of development, by an unknown mechanism \citep{Bullock.Ish-Horowicz2001,Spirov.etal2009,Fahmy.etal2014}.

\smallskip
\textbf{Approach} In whole embryo imaging, variability can arise during tissue fixation and staining with fluorophores, as well as from differences in microscope settings (gain and offset) between measurements of different batches of embryos on different days. Here, we discuss features of the data extraction which are insensitive to such experimental variation.

The aim of our approach is to create a model for basal and apical profiles (see Figure~\ref{fig3}B) with \bcd{} gradients, estimate the model
parameters and show that they can help to obtain biological results; in particular, to compare different ages in the embryo development.
We show an example of how data extracted and modelled by this technique can provide new biological insights into \bcd{} RNA gradient formation.

The novelty of the approach consists in consideration of the model parameters, which do not depend on linear transformation of the data and thereby on
the non-specific background signal and the microscope settings. It is very important, since otherwise the comparison results can be caused by the experiments conditions, not by the biology reasons.

\smallskip
\textbf{Model} A two-exponential fit of a Bcd protein profile can be well approximated by a single exponential plus a nearly-constant background \citep{Houchmandzadeh.etal2002, Alexandrov.etal2008}. In contrast, while some \bcd{} RNA profiles show such characteristics, many others, especially at early stages, show a much sharper exponential drop in the anterior, plus a constant or even posteriorly-rising component through the rest of the embryo (Figure~\ref{fig4}). The transition between components can be readily visible in RNA patterns (and not in protein), as a `kink' around the 20--30\% egg length (\%EL) position. These different components suggest multiple scales (or mechanisms) in the posterior-ward transport of \bcd{} RNA.

\smallskip
\textbf{Technique}
We previously applied a signal extraction technique based on Singular Spectrum Analysis (SSA) to quantify Bcd protein gradients \citep{Alexandrov.etal2008}.
This demonstrated that SSA could reliably and automatically extract AP Bcd protein gradients. These were the sum of two exponentials, one with a significant decay constant (strong curvature) and one of nearly linear form, capturing the non-specific background signal. Here, we adapt the SSA technique to the more complex cases of \bcd{} RNA gradients, validating the reliability and effectiveness of the approach. SSA itself is used for signal extraction, and the SSA-related method ESPRIT \citep{Roy.Kailath1989, Golyandina.Zhigljavsky2012} is used for the estimation of signal parameters.

SSA techniques have proven to be robust to signal extraction from data with substantial experimental variability and intrinsic noise \citep{Golyandina.etal2001,Alexandrov.etal2008,Alonso.etal2005,Golyandina.etal2012,Golyandina.Zhigljavsky2012}.
The use of SSA for extraction of signals in gene expression data was  in \citet{Spirov.etal2012,Golyandina.etal2012, Shlemov.etal2015,Shlemov.etal2015a}.

\begin{figure}[!htb]
\begin{center}
\textbf{A}\includegraphics[width=5in]{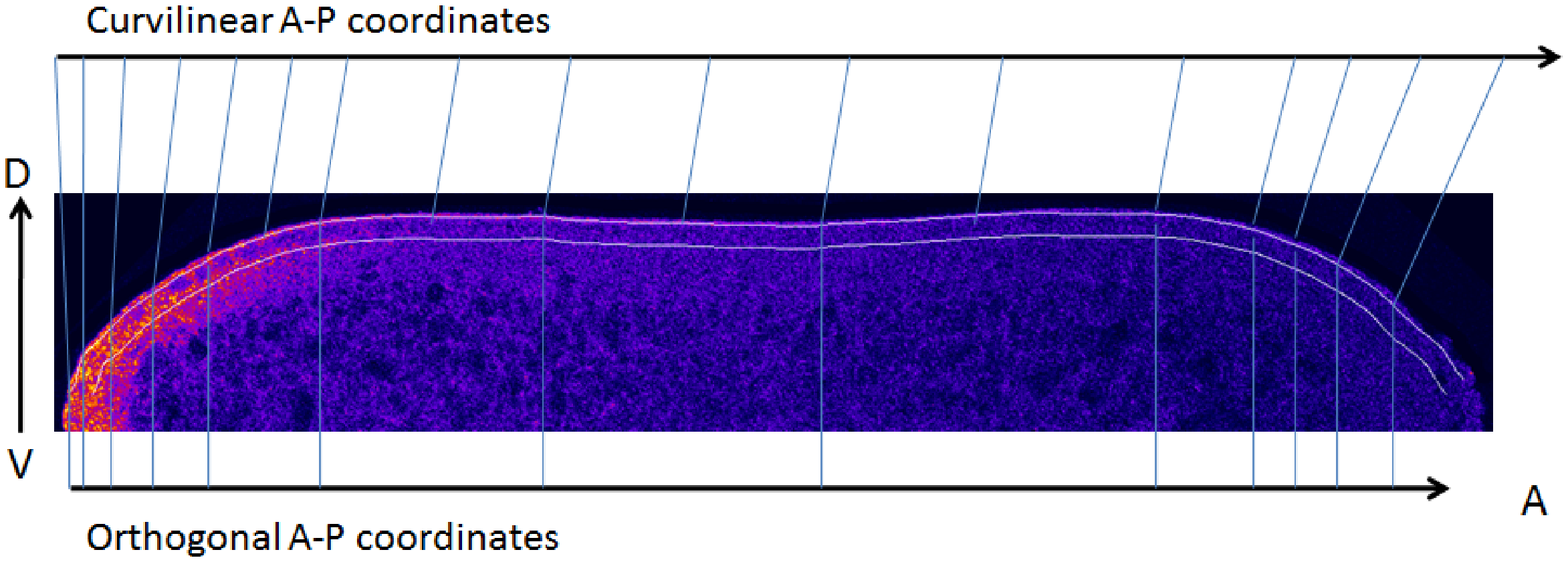} \\
\textbf{B}\includegraphics[width=4in]{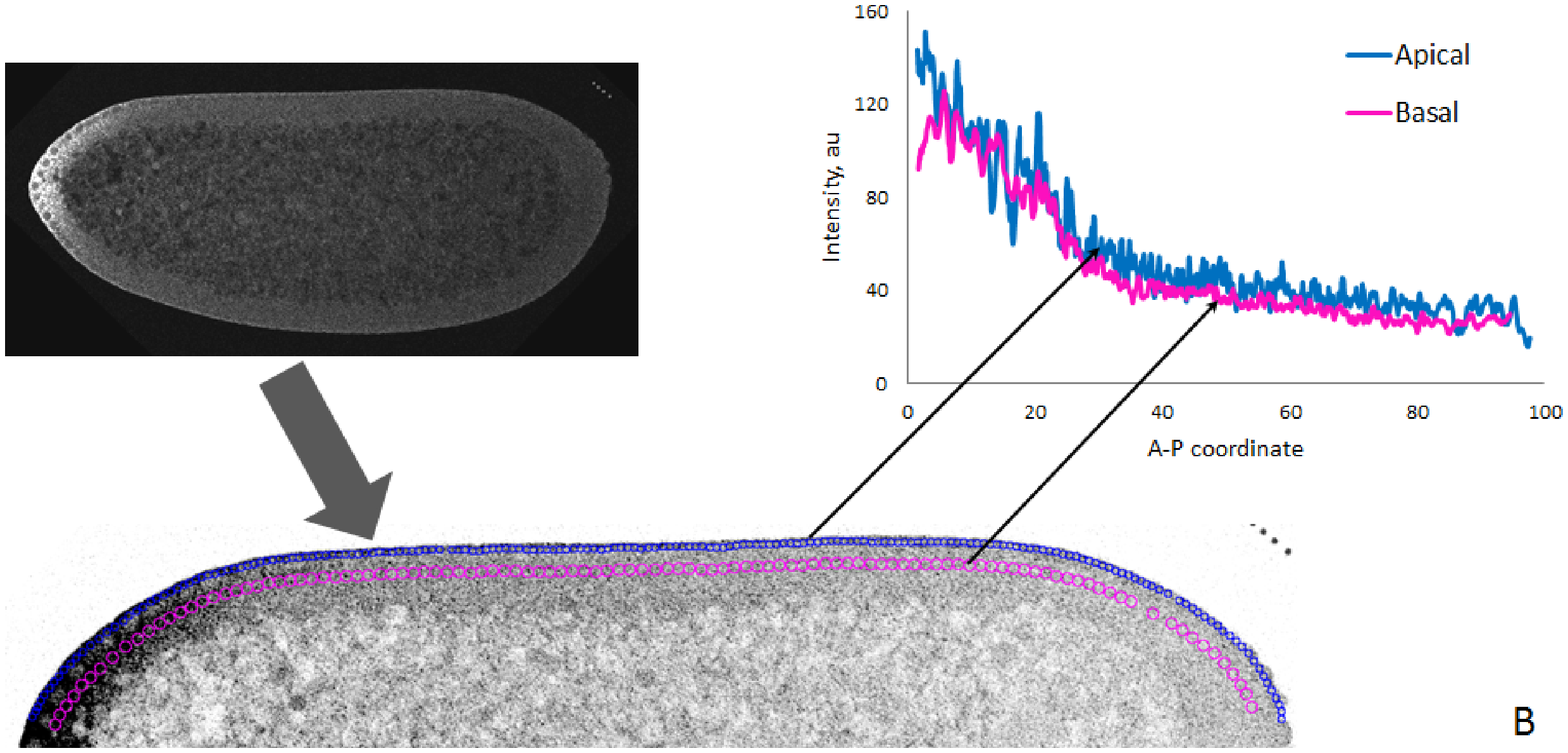}
\caption{Preparation of data for quantitative analysis of sagittal images by 1D Singular Spectrum Analysis (SSA).  \textbf{A}. Fluorescence intensity is proportional to the concentration of \bcd{} RNA. The gradient in \bcd{} mRNA is chiefly in the head-to-tail, AP, direction (left to right), but DV variation (top-to-bottom coordinate) can be seen, as well as variation by depth in the embryo (BA direction). For transport and patterning along the surface of the embryo, the natural coordinates are curvilinear. For extraction of the head-to-tail gradient patterning, the curvilinear coordinates are well approximated by a projection onto the AP axis (see text). \textbf{B}. For quantification of the AP gradient and BA differences, we sample data from an apical layer above the cortical nuclei and from a basal layer below the cortical nuclei, using chains of overlapping regions of interest (ROIs). Data from each layer is analyzed independently with 1D SSA. Each layer can then be plotted as intensity vs. AP position (right inset).}
\label{fig3}
\end{center}
\end{figure}

\begin{figure}[!htb]
\begin{center}
\textbf{A}\includegraphics[width=2.5in, height=2in]{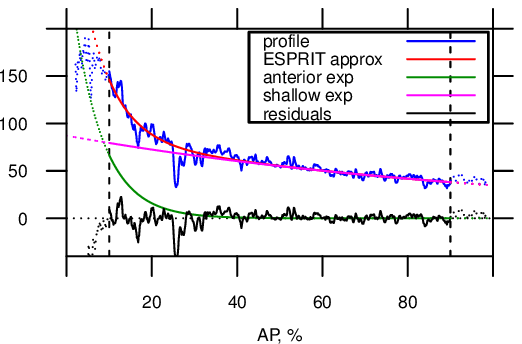}
\textbf{B}\includegraphics[width=2.5in, height=2in]{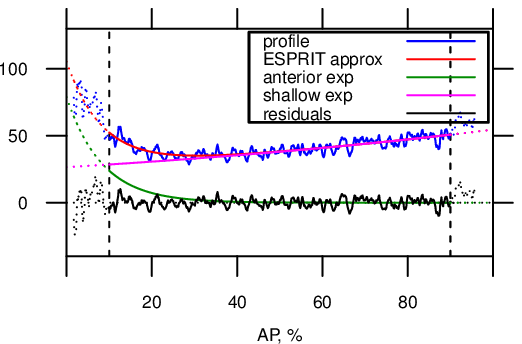} \\
\textbf{C}\includegraphics[width=2.5in, height=2in]{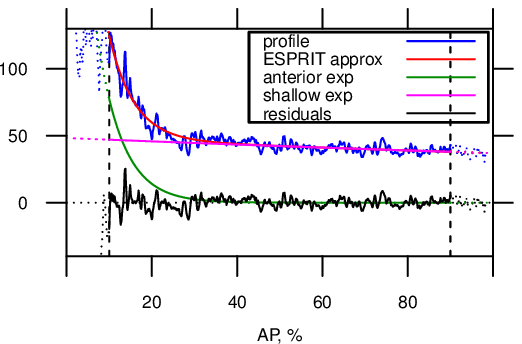}
\textbf{D}\includegraphics[width=2.5in, height=2in]{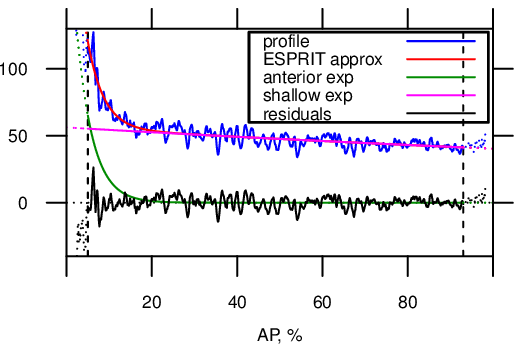}
\caption{ Representative examples of AP profiles of  \bcd{} mRNA, illustrating the variety of cases and efficacy of the modeling approach.
Blue is the original data, red is the ESPRIT fit, the sum of two exponentials (green and magenta, individually).
\textbf{A}. An early nuclear cleavage cycle 14A (nc14) embryo with a typical broad anterior exponential (green) and shallow 2nd component (magenta) extending throughout the embryo (Cf \cite{Spirov.etal2009}).
 \textbf{B}. A \bcd{} mRNA profile in which the 2nd, trunk, component rises towards the posterior (i.e. has a positive exponential rate).
 \textbf{C}. A case with a nearly flat 2nd component (representing the mRNA signal posterior of 25\%EL).
 \textbf{D}. An embryo with a very sharp anterior (1st component) exponential, dropping to low values by 10\%EL. }
\label{fig4}
\end{center}
\end{figure}

\clearpage
\section{Methods}
\subsection{Two-exponential modeling}

We fit the following two-exponential function (of AP distance, x) to \bcd{} mRNA data, to capture the distinct two-component pattern of most \bcd{} RNA gradients (with the `kink', commonly observed at ~20--30 \%EL):

 \begin{equation}
 \label{eq:signal}
s(x)=C_{\textrm{anterior}} e^{\alpha_{\textrm{anterior}} x}+C_\textrm{shallow} e^{\alpha_\textrm{shallow} x},
\end{equation}
or
\[s(x)=C_{\textrm{anterior}} \lambda_{\textrm{anterior}}^{x} +C_\textrm{shallow} \lambda_\textrm{shallow}^{x},\]
for $\lambda = e^\alpha$. The two components, $\textrm {anterior}$ (for the sharp, quickly decaying pattern in the anterior) and $\textrm {shallow}$ (for the more constant component in the mid- and posterior embryo), each have two parameters - an amplitude $C$, and a decay $\lambda$. The anterior exponential is always decreasing and therefore $\lambda_\textrm{anterior}<1$, $\alpha_\textrm{anterior}<0$; while the shallow exponential can be decreasing or increasing (Figure~\ref{fig3}). In biological terms, $C$ for $\lambda<1$ represents the maximum concentration of the component, and $\lambda$ represents the rate at which the component decreases (or increases) along the AP coordinate.

One-exponential plus constant background is a special case of equation \eqref{eq:signal}, with
$\lambda_\textrm{shallow} =1$ (or $\alpha_\textrm{shallow}=0$). Use of \eqref{eq:signal} does not require two strong (nonzero $\alpha$) exponentials in the signal (pattern). In the case of the model commonly applied to the Bcd protein gradient, the first exponential describes the signal and the second exponential describes the non-specific background signal and the offset of the microscope.

Raw image data is likely of the form $s(x)+\varepsilon (x)$, where $\varepsilon (x)$ represents ``noise'', i.e. non-regular oscillations with zero mean.

\subsubsection{Model characteristics independent of the microscopy gain/offset and background}

To remove effects from variability in microscope settings (gain and offset) and the unknown form of non-specific background  \citep{Houchmandzadeh.etal2002, Myasnikova.etal2005,Holloway.etal2006}, gradient characteristics can be used which do not change under a linear transformation of the gradient.

That is, if each profile (gradient) can be represented by the linear transformation:
 \begin{equation} \label{0.1)}
f(x)=A(s(x)+\varepsilon (x))+B,
\end{equation}
where A and B represent an unknown scaling and an unknown offset, respectively. A and B are likely to differ between embryos (with different staining conditions, microscope settings, etc.), but when we take basal and apical traces within each embryo image, we assume that the A and B are constant within a single embryo. To compare data between embryos, we take advantage of the independence of the following profile characteristics from linear transformations, i.e. independence from A and B values.

\begin{itemize}
\item  the ratio between the anterior gradient pre-exponential coefficients for the apical and basal profiles
$C^\textrm{ab} = \ln( C_\textrm{anterior}^\textrm{apical}/ C_\textrm{anterior}^\textrm{basal})$
;
\item
 the following ratio for the shallow component\\ $C_{\textrm{shallow}}^{({\rm apical})} \alpha _{\textrm{shallow}}^{({\rm apical})} /(C_{\textrm{shallow}}^{({\rm basal})} \alpha _{\textrm{shallow}}^{({\rm basal})})$;
\item  indicators of non-increase in the shallow components $\lambda_\textrm{shallow}^\textrm{apical}\le 1$ and $\lambda_\textrm{shallow}^\textrm{basal}\le 1$;
\item  in addition, the AP position at which the anterior components become almost zero $AP0_{{\rm anterior}}^{({\rm apical})} $ and $AP0_{{\rm anterior}}^{({\rm basal})} $.
\end{itemize}

These relations underlie the quantitative conclusions in this paper. We also use these relations to screen for non-typical embryos, which aids in following the development of the \bcd{} RNA gradient over time and for studying apical-basal differences.

\paragraph{Mathematical details}
\label{sec:math}

We will use index 1 for {\textrm{anterior}} and index 2 for {\textrm{shallow}}.
The signal \eqref{eq:signal} has characteristics which approximately satisfy independency from linear transformations if the second (shallow-gradient) exponential rate is small enough ($\lambda _{2} {\rm \approxeq }1$, $\alpha _{2} {\rm \approxeq }0$) and can therefore be approximated by a linear function. This is a reasonable assumption for the \bcd{} mRNA data, giving
\begin{gather*}
A(C_{1} \exp (\alpha _{1} x)+C_{2} \exp (\alpha _{2} x))+B\approx \\
AC_{1} \exp (\alpha _{1} x)+AC_{2} (1+\alpha _{2} x)+B\approx
 \tilde{C}_{1} \exp (\tilde{\alpha }_{1} x)+\tilde{C}_{2} \exp (\tilde{\alpha }_{2} x),
\end{gather*}
where
\begin{gather*}
\tilde{C}_{1} =AC_{1} ,\quad \tilde{\alpha }_{1} =\alpha _{1} ,\quad \tilde{C}_{2} =AC_{2} +B,\quad
 \tilde{\alpha }_{2} =AC_{2} \alpha _{2} /(AC_{2} +B).
\end{gather*}
Thus, the following characteristics of the profiles can be considered as almost independent of a linear transformation of the intensities, i.e., of $A$ and $B$:
\begin{gather} \label{ZEqnNum155807}
\alpha _{1}^{({\rm apical})} ,\qquad  C_{2}^{({\rm apical})} \alpha _{2}^{({\rm apical})} /C_{1}^{({\rm apical})} ,\\
\label{ZEqnNum701878}
\alpha _{1}^{({\rm basal})} ,\qquad C_{2}^{({\rm basal})} \alpha _{2}^{({\rm basal})} /C_{1}^{({\rm basal})} ,\\
C_{1}^{({\rm apical})} /C_{1}^{({\rm basal})},\qquad
\label{ZEqnNum733187}
C_{2}^{({\rm apical})} \alpha _{2}^{({\rm apical})} /(C_{2}^{({\rm basal})} \alpha _{2}^{({\rm basal})} ),
\end{gather}
where \eqref{ZEqnNum155807} are characteristics of apical profiles, \eqref{ZEqnNum701878} are characteristics of basal profiles, and characteristics \eqref{ZEqnNum733187} show relations between apical and basal profiles.
If $C_{2} >0$ and $AC_{2} +B>0$ (true, generally, for the \bcd{} RNA data),  the sign of $\alpha _{2} $ is not affected by a linear transformation and the second exponential can be either increasing or decreasing.

\subsubsection{Estimation of the two-exponential model parameters}

  We use the subspace-based method ESPRIT, motivated by the success of SSA (also a subspace-based method) in smoothing one-dimensional gene profiles from \Drosophila{} embryos \citep{Alonso.etal2005, Golyandina.etal2012}. On profiles from different genes, the method proved to be robust to high noise and to variations in embryo characteristics.

The mathematical details of ESPRIT can be found in the Supplementary material. We use the method to estimate the exponential decays in \eqref{eq:signal}: $\lambda_\textrm{anterior}^\textrm{(apical)}$, $\lambda_\textrm{anterior}^\textrm{(basal)}$, $\lambda_\textrm{shallow}^\textrm{(apical)}$ and $\lambda_\textrm{shallow}^\textrm{(basal)}$. The estimation of the coefficients $C_\textrm{anterior}^\textrm{(apical)}$, $C_\textrm{anterior}^\textrm{(basal)}$, $C_\textrm{shallow}^\textrm{(apical)}$ and $C_\textrm{shallow}^\textrm{(basal)}$ are then found by conventional least-squares, since the model \eqref{eq:signal} is linear in these parameters.

Since the first exponential is expected to be rapidly decreasing ($\lambda_{\textrm{anterior}} <1$) and the second exponential is expected to be close to constant ($\lambda_{\textrm{shallow}} $ near 1), we reorder the ESPRIT estimates accordingly.

\subsection{Data}
\label{sec:data}
\subsubsection{ FISH and data acquisition}

Fluorescent in situ hybridization (FISH) for \bcd{} mRNA is as described in \citet{Spirov.etal2009}, see the Supplementary materials for more details. Computational tools to process midsagittal images are described in the Supplementary Materials too.
Our dataset consists of images of about 160 embryos, ranging in stage from unfertilized eggs (not analyzed) to early nuclear cleavage cycle 14A (nc14, same dataset as in \citet{Spirov.etal2009}). In the current study, we analyzed 124 embryos. These were divided into three developmental stages, based on preliminary analysis and biological considerations: Cleavage, or pre-blastoderm (nc1--nc9); Syncytial Blastoderm (nc10--nc13); and Cellularizing Blastoderm (nc14A). The Cleavage stage is long, lasting about 80 min (at room temperature), and has highly variable bcd mRNA gradients. For more detailed analysis, we subdivided Cleavage into two sub-groups: Early (nc1--nc8) and Late (nc9). The Syncytial Blastoderm stage spans about 45 min, and this could be subdivided into two sub-groups: nc10--nc12 and nc13). The last stage, early nc14A, is short (15--20 min.), but highly variable and dynamic. Careful visual inspection allowed us to divide the nc14A embryos on three sub-groups: early, mid and late \citep{Spirov.etal2009}.

\subsubsection{Construction of 1D profiles}
Raw data from the confocal microscope consists of mRNA intensities per a small circular area with 2D spatial coordinates. After selecting the regions of interest (ROI chains), two techniques were tested for converting the data into 1D AP profiles.  The first (and simplest) technique projects intensities onto an AP axis orthogonal to the DV axis by discarding the DV component of the coordinate (Figure ~\ref{fig3}A). This has been used by many groups, see for example \citet{Surkova.etal2008,Houchmandzadeh.etal2002}). The second technique preserves the natural curvilinear coordinates of the embryo, with distance between ROIs calculated by
 $d^{2} (i)=({\rm AP}(i+1)-{\rm AP}(i))^{2} +({\rm DV}(i+1)-{\rm DV}(i))^{2}$.
 Cumulative distances are then normalized by dividing by the sum of $d(i)$.

 Regardless of the technique (AP or curvilinear coordinates), the 1D coordinates obtained are not equidistant. Linear interpolation was used to create equidistant points of a given spatial step. A step 0.08--0.1\%EL was chosen to generate approximately equal numbers of points for the two techniques.
 These results obtained by means of AP coordinates appear to be more precise than that obtained by curvilinear coordinates, see the Supplementary material
 for comparison. Therefore, we can consider only AP coordinates in the paper.

\section{Results and discussion}

\subsection{Model application}

 Figure~\ref{fig4} demonstrates a set of examples to illustrate the variety of profiles which can be fit by the two-exponential model. These include the typical profiles considered in \citet{Spirov.etal2009}, with a rapidly decreasing anterior gradient and slowly decreasing gradient to the posterior (Figure~\ref{fig4}A), and profiles with increasing (Figure~\ref{fig4}B) or flat (Figure~\ref{fig4}C) posterior gradients.

Data is generally too biased and noisy from the terminal regions of the embryo: 0--10 \%EL and 90--100 \%EL(Cf \citep{Surkova.etal2008,Houchmandzadeh.etal2002}, see Figure~\ref{fig4}). Processing and analyzing data from 10--90 \%EL is sufficient for extracting \bcd{} RNA profiles from nearly all embryos older than nc6. For very early embryos (CleavageEarly stage), gradients have just begun to form from initial terminal locales; in these cases, we process from 5--93 \%EL Figure~\ref{fig4}D.
Figure~\ref{fig4} shows that the two-exponential model suits different types of data very well.

Typically, embryos have a decreasing anterior exponential component and decreasing or close-to-constant shallow posterior gradients (both for the apical and basal profiles). We call these Type 1 (typical) embryos. Some embryos, however, show a posteriorly-increasing shallow gradient for either apical or basal profiles. We call these Type 2 (atypical) embryos. Type 2 profiles are common early in development (Cleavage) and uncommon in later stages.
Here, we focus on Type 1 embryos, which represent the majority of the dataset.

Detection of Type 1 can be performed by means of exponential rates of the shallow exponents:
(A) $\lambda_\textrm{shallow}^\textrm{(apical)}<1.002$, $\lambda_\textrm{shallow}^\textrm{(basal)}<1.002$. This condition screens for shallow profiles (both basal and apical) which do not increase towards the posterior  (1.002 is used for 1, to account for estimation errors).

\subsection{Model validation}
\label{sec:selection}
 Even within one developmental stage, the shape of mRNA profiles from embryo to embryo is highly variable. This makes construction of a prototype profile challenging, and complicates understanding of the underlying biological mechanisms.
 Fortunately, the variability is mostly due to a minority of embryos, and these can be detected using the two-exponential model. Removal of such embryos reduces the variability significantly.

 The outliers can be found by standard tools basing on 2D scatterplots of the estimated parameters. It appears that the outliers can be removed by the conditions data (B) $AP0_{{\rm anterior}}^\textrm{(apical)} \le 120$, $AP0_{{\rm anterior}}^{({\rm basal})} \le 120$ and
 (C) $C^{ab}>-1$.
  These constraints have natural biological interpretations: (B) screens out embryos whose anterior gradient (apical or basal) decreases too slowly and cannot be distinguished from the shallow gradient (i.e. the embryo is not described by the two-exponential model); and (C) screens for bad estimates of the apical vs. basal intensities (see definition of $C^{ab}$ in the Introduction). (A), (B), (C) are robust to small changes in constraints thresholds (results not shown).

  Figure 2 in the Supplementary material shows scatterplots of $\lambda_\textrm{anterior}^\textrm{(apical)}$,  $\lambda_\textrm{anterior}^\textrm{(basal)}$, $\lambda_\textrm{shallow}^\textrm{(apical)}$ and $\lambda_\textrm{shallow}^\textrm{(basal)}$ before and after application of the constraints. Most outlier embryos were filtered out, making the distribution of profile parameters more homogeneous.

88 embryos satisfy conditions (A)--(C), from the complete dataset of 124 embryos; the analysis in the rest of this paper is on these 88 embryos.
For these 88 embryos, it was checked that the systematic errors in the model is negligible relative to the residual noise or to the profile itself.
Thus, we conclude that the profiles suit the considered model with sufficient accuracy (see the Supplementary material for details).

\subsection{Model efficacy for finding trends in developmental biology}

  The parameters from the two-exponential fits are quite variable, both within and between developmental stages (\ref{fig:groups_diff}A), as expected from the observed variability in profiles (Section~\ref{sec:intro}).

Though the large variability and small sample size do not allow for statistically significant conclusions for all comparisons, several observations can be made. In particular, CleavageEarly has the largest average anterior exponential decay constant of any developmental stage (i.e. the steepest profile). This difference is statistically significant (t-test), but could be rendered insignificant by moderate changes in just one of the six embryos. We therefore combine groups to obtain 3 age groups (from 7) with larger sample sizes: (1) Cleavage, $n = 19$; (2) nc10--nc13, $n = 23$; (3) nc14, $n = 46$. Figure~\ref{fig:groups_diff}B shows that these larger groups have more distinct clustering, with distinct means.

\begin{figure}[!htb]
\begin{center}
A.\includegraphics[width=3in]{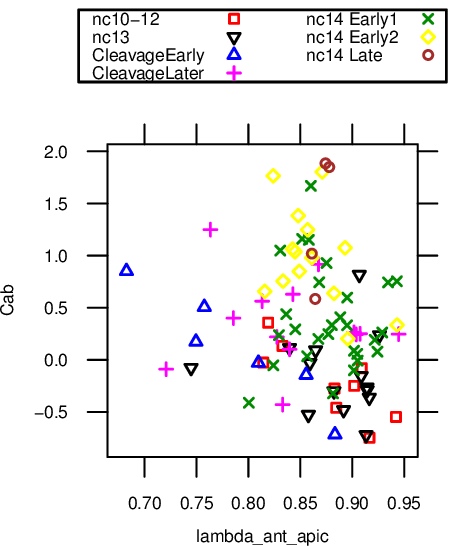}
B.\includegraphics[width=3in]{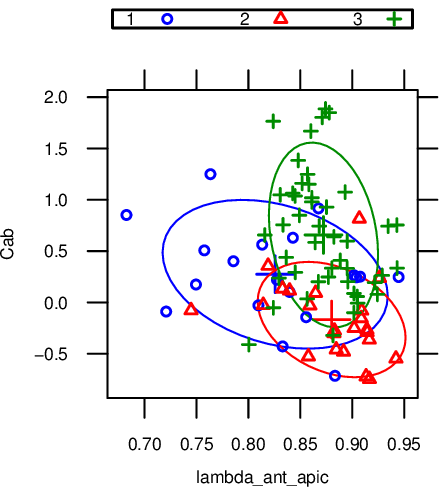}
\caption{Pre-exponential factor $C^{ab}$ vs. anterior gradient $\lambda_\textrm{anterior}^\textrm{(apical)}$. \textbf{A}. 7 (marked) developmental stages (Section~\ref{sec:data}): Two-exponential parameters show large variability between and within the developmental stages. \textbf{B}. 3 combined groups (see key): difference in parameter values, with 90\% confidence ellipsoid.}
\label{fig:groups_diff}
\end{center}
\end{figure}

Table~\ref{tab:combined_means} shows the average values for $\lambda_\textrm{anterior}^\textrm{(apical)}$ and $C^{ab}$ with their 90\% confidence intervals.

\begin{table}[htb]
\begin{center}
\caption{Combined groups: means and 90\% confidence intervals for main characteristics of apical and basal profiles for 3 groups.}
\label{tab:combined_means}
\begin{tabular}{|l||r|r|r|} \hline
 & $\lambda_\textrm{anterior}^\textrm{apical}$ & lower bound & upper bound\\ \hline
Cleavage    & 0.825 & 0.798 & 0.853 \\ \hline
nc10-13  & 0.880 & 0.863 & 0.896 \\ \hline
nc14     & 0.872 & 0.863 & 0.880 \\ \hline \hline
& $C^\textrm{ab}$ & lower  bound& upper bound\\ \hline
Cleavage    & 0.274 & 0.089 & 0.459 \\ \hline
nc10-13  & -0.167 & -0.296 & -0.037 \\ \hline
nc14     & 0.657 & 0.511 & 0.803 \\ \hline
\end{tabular}
\end{center}
\end{table}

One-way ANOVA (both parametric and non-parametric (Kruskal-Wallis)) confirms that both $\lambda_\textrm{anterior}^\textrm{(apical)}$ and $C^{ab}$ significantly differ between the groups at the 5\% level.
Post-hoc comparisons show that $C^{ab}$ (the logarithm of the ratio between the apical and basal anterior gradients at 10 \%EL) is significantly different between all three groups; while the exponential decay rate of the anterior gradient is significantly larger ($\lambda$ is smaller) only for the Cleavage group.

\subsubsection{Potentials of the approach}
In section~\ref{sec:selection}, we screened embryos into Type 1 using condition (A), i.e. non posteriorly-increasing profiles.
 We can apply the suggested approach to embryos of Type 2 with posteriorly-increasing profiles (see Figure~\ref{fig4}D). Moreover, the model can be extended to three exponentials (see Figure~\ref{fig:3exp}). With the extension of SSA to fit a 3-exponential model, these sorts of patterns can be readily analyzed by the current approach, broadening the use of the technique to allow for the comparison of patterns from different genes (e.g., consider Stau protein \citep{Spirov.etal2009}, which has a sharp rise in the vicinity of the posterior pole).

\begin{figure}[!htb]
\begin{center}
\includegraphics[width=4in, height=3in]{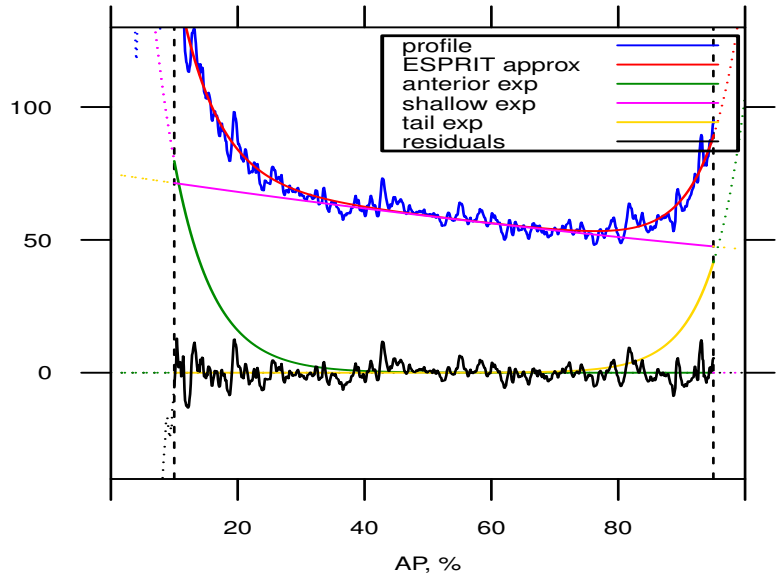}
\caption{AP profile of the Stau protein (cf. \citet[Fig.6]{Spirov.etal2009}). The raw data is in blue, the 3-exponential model is in red: anterior exponent 1, green; shallow exponent 2, magenta; posterior exponent 3, yellow; residuals, black.}
\label{fig:3exp}
\end{center}
\end{figure}

The approach presented here is likely to be an effective tool for quantifying other spatial gradients in developmental biology, which could aid in revealing new features in the patterning dynamics and regulation of critical developmental events, especially where there are large dynamic changes and high variability --- i.e. in cases where it is difficult to construct a reference or prototype profile. Examples include the Dorsal gradient in DV \Drosophila{} patterning \citep{Kanodia.etal2012,Kanodia.etal2011,Reeves.etal2012} and retinoic acid in vertebrate embryos \citep{Schilling.etal2012}.

\section{Conclusions}

 The new mathematical model described here enables the study of substantial quantitative problems in \bcd{} mRNA gradient formation, including quantification of the between-embryo variability of the gradient; the filtering of atypical gradients; and the classification of embryos on the basis of quantitative gradient parameters. We are using these abilities to quantitatively study the dynamics of \bcd{} mRNA profiles at very early stages of development. Finally, we can also now use the new mRNA gradient model to compare mRNA patterning with the Bcd protein gradient, previously analyzed in \citet{Alexandrov.etal2008}.

\section*{Acknowledgement}
This work has been supported by U.S. NIH grant R01-GM072022 and
the Russian Foundation for Basic Research grants 15-04-06480 and 16-04-00821.

\bibliographystyle{abbrvnat}

\clearpage
\section*{Supplementary materials}

\appendix

\section{Data}

\subsection{FISH and data acquisition}
Images ($1024\times 1024$  pixels, 8 bit) were taken using confocal microscopy \cite{Spirov.etal2009}. Images were acquired through whole embryo stacks, and suitable mid-sagittal slices were selected to eliminate unnecessary geometric distortion. Using the raw data directly from the confocal microscope, intensities were measured by sliding an area perpendicular to the embryo edge, similar to \cite{Houchmandzadeh.etal2002}, but with our own algorithms, scripts and tools (below). One curve followed the dorsal apical periplasm, while the other curve followed the dorsal basal periplasm.

\subsection{Computational Tools to Process Images}
Our tools consisted of a set of plug-ins for ImageJ software (W. Rasband, NIH, USA) and scripts in Delphi (for Windows) or GnuPascal (for Linux). After raw image rotation and cropping, the software is used to find the image contour (dorsal or ventral edge of embryo). This contour is used to find a series of curvilinear profiles (lines) running beneath and in parallel to the contour. The two main (apical and basal) profiles were chosen by visual inspection. Local intensity data was collected along these profiles. A small circular window of a given radius R (in pixels), or Region Of Interest (ROI), is centered on the profile. The ROI is slid in steps of one pixel along the profile. At each step, the intensity is measured and averaged over the ROI and saved. This method of measuring overlapping areas of averaged intensities served as a first step in de-noising the (noisy) FISH images. (A second step of de-noising was done with SSA, see below). For apical profiles, a radius of 3 pixels was used for the ROI, covering a thin layer of apical periplasm between the nuclear membrane and plasma membrane along the dorsal axis. For the basal periplasm, two radii, R = 5 and R = 12 pixels, were tested (the basal periplasm is substantially wider than the apical periplasm). R = 5 is sufficient to collect the representative data. To the best of our knowledge, all prior work on such \Drosophila{} data has involved a projection of the natural (ellipsoidal) surface curvilinear coordinates to the AP axis (running down the center of the embryo; (\cite{Gregor.etal2005,Gregor.etal2007,Gregor.etal2008,Houchmandzadeh.etal2002,Little.etal2011,Cheung.etal2011,Cheung.etal2014}). The distortion of patterns at the very tip of an embryo for such a projection could be substantial. Therefore, we tested the present results on both the natural curvilinear coordinates and on the AP projection.

\section{Mathematical details of ESPRIT}
 \label{sec:esprit}

Here, we describe the ESPRIT method (specifically LS-ESPRIT \cite{Roy.Kailath1989},\cite[Section 3.8.2]{Golyandina.Zhigljavsky2012}) applied to a sequence of observations $x_{1} ,x_{2} ,\ldots ,x_{N} $, where
\begin{equation}
\label{eq:signal1}
   x_n = s_n + \varepsilon_n,\qquad s_n=C_{1} \exp (\alpha _{1} n)+C_{2} \exp (\alpha _{2} n),
\end{equation}
$\varepsilon_n$ is a noise.

  Fix the signal rank $r$ (number of exponentials in our case) and choose a window length $r+1\le L\le N-r$. We chose $L\approx N/2$ to get better separability of the signal from noise, see \cite{Golyandina2010}, and have $r=2$.
 The first step consists in the construction of the trajectory matrix $\mathbf{X}$ from the column vectors $X_{i} =(x_{i} ,\ldots ,x_{i+L-1} )^{{\rm T}} $, $i=1,\ldots ,K=N-L+1$: $\mathbf{X}=[X_{1} :\cdots :X_{L} ]$. ESPRIT is based on the Singular Value Decomposition (SVD) of the matrix $\mathbf{X}$. Let $\mathbf{U}=[U_{1} :\cdots :U_{r} ]$ be the matrix consisting of the $r$ leading left singular vectors of $\mathbf{X}$. Denote by $\underline{\mathbf{U}}$ the matrix $\mathbf{U}$ without the last row and by $\overline{\mathbf{U}}$ the matrix $\mathbf{U}$ without the first row. Consider the $(r\times r)$-matrix $\Lambda =\underline{\mathbf{U}}^{-} \overline{\mathbf{U}}$, where $\mathbf{A}^{-} $ stands for pseudo-inverse of $\mathbf{A}$.
 The eigenvalues of $\Lambda$ provide the ESPRIT-estimates $\widetilde\lambda_{i}$ of $\lambda_{1} $ and $\lambda_{2} $, where
 $\lambda_{i} =\exp (\alpha_{i})$, see \eqref{eq:signal1}.
 Coefficients $C_{i} $ in \eqref{eq:signal1} can be found by means of the ordinary least-squares method in the linear model
 $x_n = C_1 z_n^{(1)} + C_2 z_n^{(2)} + \varepsilon_n$, where $z_n^{(i)} = \lambda_{i}^n$.

\clearpage
\section{Data processing}

\subsection{Filtering of embryos}
 Figure~\ref{fig:matrixplot}  shows scatterplots of $\lambda_\textrm{anterior}^\textrm{(apical)}$,  $\lambda_\textrm{anterior}^\textrm{(basal)}$, $\lambda_\textrm{shallow}^\textrm{(apical)}$ and $\lambda_\textrm{shallow}^\textrm{(basal)}$ before and after application of the constraints. Most outlier embryos were filtered out, making the distribution of profile parameters more homogeneous.

 We consider embryos with non-increasing shallow exponential: (A) $\lambda_\textrm{shallow}^\textrm{(apical)}<1.002$, $\lambda_\textrm{shallow}^\textrm{(basal)}<1.002$. The outliers can be found by standard tools basing on 2D scatterplots of the estimated parameters. It appears that the outliers can be removed by the conditions data (B) $AP0_{{\rm anterior}}^\textrm{(apical)} \le 120$, $AP0_{{\rm anterior}}^{({\rm basal})} \le 120$ and
 (C) $C^{ab}>-1$.

Table~\ref{tab:valid} shows the proportion of embryos, by stage, satisfying conditions (A), (B), and (C). The number of embryos excluded is highest for Cleavage stage; exclusion here is chiefly by criterion (A), reflecting that early stage embryos are more likely to have posteriorly-increasing shallow gradients than later stage embryos (33\% for Cleavage; 17\% for nc10-13; 7\% for nc14. (For embryos not excluded by criteria (B) and (C), the proportion of the embryos with posteriorly-increasing shallow gradient was: 17\% Cleavage; 8\% nc10-13; 0\% nc14.)

\begin{table}[!htb]
\begin{center}
\caption{Proportion of embryos satisfying criteria (A), (B) and (C) individually and combined, from the whole dataset of embryos.}
\label{tab:valid}
\begin{tabular}{|l|r|r|r|r|r|} \hline
group & N &(A) & (B) & (C) & (A),(B),(C)\\ \hline
Cleavage	&19&66\%	&76\%	&79\%	&50\%\\ \hline
10-13 cycle	&23&83\%	&90\%	&90\%	&77\%\\ \hline
14 cycle	&46&93\%	&88\%	&91\%	&82\%\\ \hline
\end{tabular}
\end{center}
\end{table}

\begin{figure}[!htb]
\begin{center}
\textbf{A}\includegraphics[width=6in]{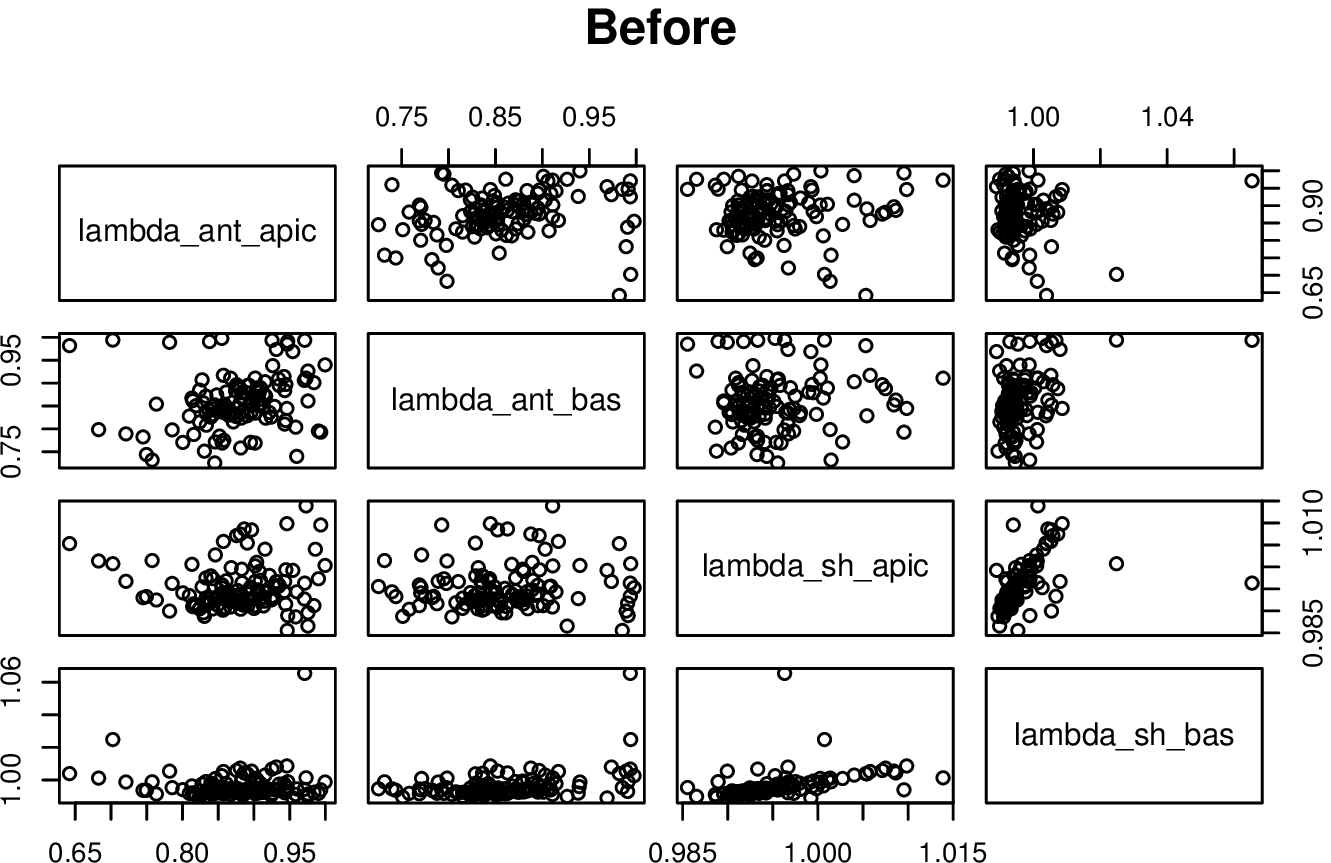} \\
\textbf{B}\includegraphics[width=6in]{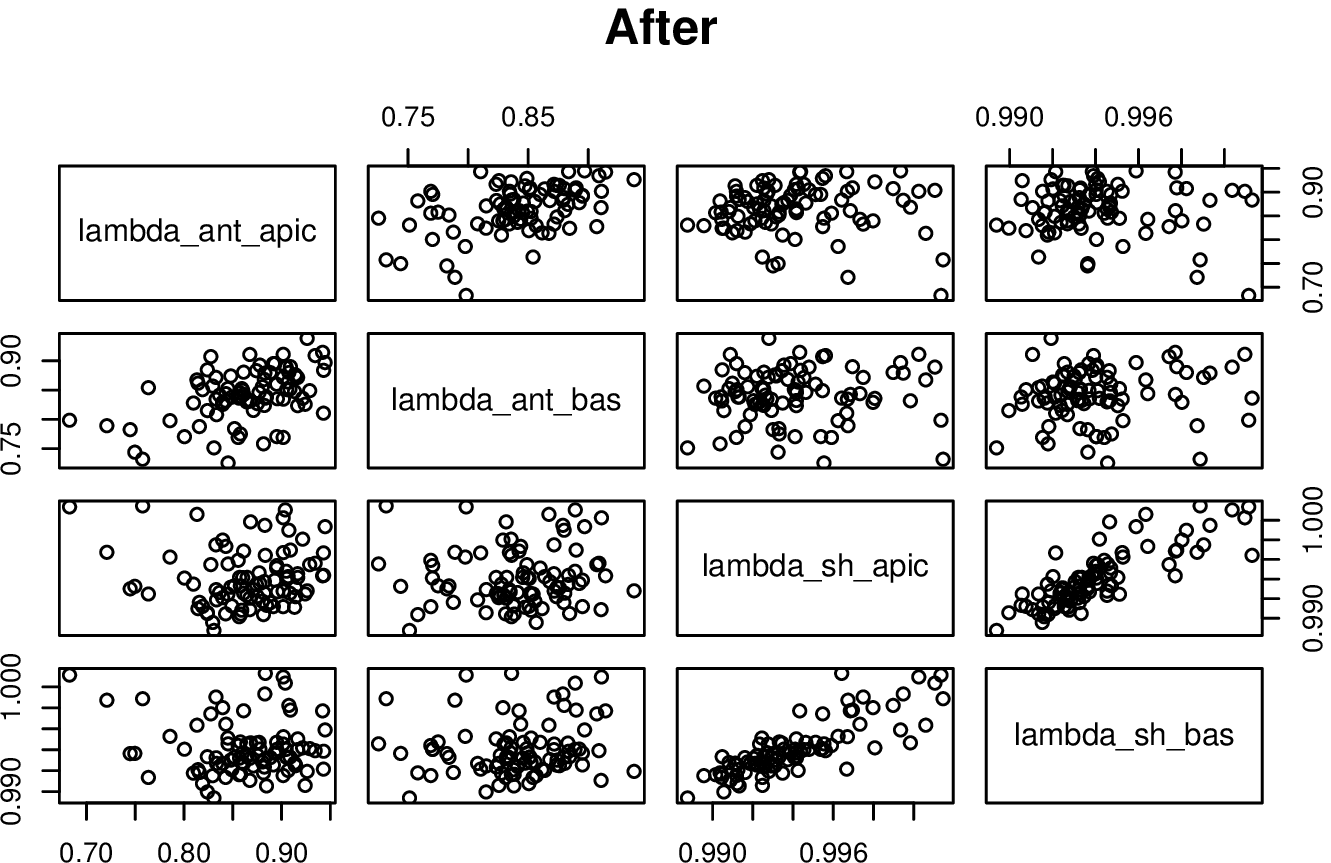}
\caption{Scatterplots of $\lambda_\textrm{anterior}^\textrm{(apical)}$, $\lambda_\textrm{anterior}^\textrm{(basal)}$, $\lambda_\textrm{shallow}^\textrm{(apical)}$, and $\lambda_\textrm{shallow}^\textrm{(basal)}$, before (A) and after (B) application of the constraints.}
\label{fig:matrixplot}
\end{center}
\end{figure}

88 embryos satisfy conditions (A)--(C), from the complete dataset of 124 embryos (Table~\ref{tab:valid}); the analysis in the rest of this paper is on these 88 embryos.

\clearpage
\subsection{The model accuracy}
\label{sec:accuracy}
 \paragraph{Systematic error. }
Here, we analyze the adequacy of the two-exponential model using an example from the Early2 sub-stage of nc14; these results are typical of the two-exponential fit. It can be seen in Figure~\ref{fig:residuals}A (black) that the noise is not homogeneous (it has changing variability); the averaged residuals (red line; Figure~\ref{fig:residuals}A) show this systematic error as a function of the AP coordinate. However, this variation is of magnitude no more than two intensity units (greatest near the inflection point, at the switch between the two exponential components), which is negligible relative to the residual noise or to the profile itself Figure~\ref{fig:residuals}B.

\begin{figure}[!htb]
\textbf{A}\includegraphics[width=3in, height=1.5in]{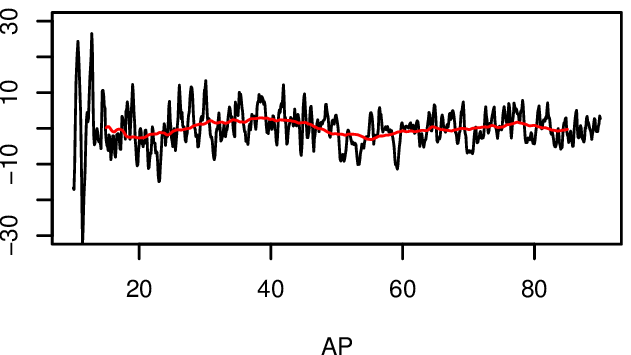}
\textbf{B}\includegraphics[width=3in, height=2in]{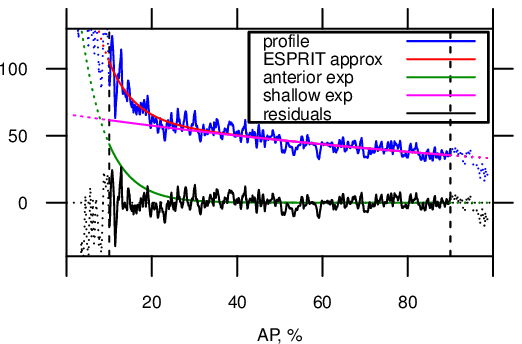}
\caption{Noise and trend for a cycle 14 embryo.  A. Residuals, showing a systematic variation. B. Model fitting to data (above) and residuals (below).}
\label{fig:residuals}
\end{figure}

Table~\ref{tab:residuals2exp} shows the root mean square error (RMSE) of residuals across developmental stages. These are always significantly smaller than the magnitude of the profile intensities. Table~\ref{tab:residuals2exp} shows that the systematic error, computed by applying a median filter of 40\%EL, never exceeds 1.0, negligible on the profile intensity range.
 (Note that both the RMSE of residuals and of systematic error components decrease with developmental age; earlier profiles show stronger noise. However, this effect is not independent of a linear transformation of profiles and therefore can be caused by different microscope settings.)
Overall, the model closely approximates the data profiles, leaving chiefly non-systematic noise in the residuals.


\subsection{Comparison with `exponential plus constant' model}

\begin{table}[!htb]
\begin{center}
\caption{Characteristics of residuals after fitting the `exponential plus constant' and the two-exponential models to the mRNA profiles.}
\label{tab:residuals2exp}
\begin{tabular}{|l|r|r|r|r|r|r|} \hline
 & \multicolumn{2}{|c|}{exp+const}  & \multicolumn{4}{|c|}{2-exp}  \\ \hline
 & \multicolumn{2}{|c|}{residual RMSE} & \multicolumn{2}{|c|}{residual RMSE} & \multicolumn{2}{|c|}{systematic RMSE} \\ \hline
 & apical & basal & apical & basal & apical & basal \\ \hline
Cleavage & 5.5 & 5.0& 4.7 & 4.7 & 1.0 & 1.0 \\ \hline
nc10--13 & 4.7 & 4.7 & 4.2 & 3.0 & 0.8 & 0.7 \\ \hline
nc14 & 5.5 & 2.6 & 4.1 & 2.0 & 0.6 & 0.4 \\ \hline
\end{tabular}
\end{center}
\end{table}

 Single-exponential-plus-constant-background models have been used in a number of studies of /Bcd{} profiles, both protein \cite{Houchmandzadeh.etal2002} and mRNA \cite{Spirov.etal2009}. Table~\ref{tab:residuals2exp} demonstrates that the two-exponential model fits the mRNA profiles better (MSE is smaller) than such exponential-plus-constant models. These results are not too surprising, since the two-exponential model is not constrained to have a flat background.

\subsection{Curvilinear coordinates}

Here, we test that the results of fitting the two-exponential model is robust to small non-linear transformations. That is, we test for differences in using AP projections (ignoring the DV coordinate) vs. curvilinear coordinates along the profile in the image ($d^{2} (i)=({\rm AP}(i+1)-{\rm AP}(i))^{2} +({\rm DV}(i+1)-{\rm DV}(i))^{2}$; cumulative distances are then normalized by dividing by the sum of $d(i)$): see Figure~\ref{fig:curvilinear}.

\begin{figure}[!htb]
\begin{center}
\textbf{A}\includegraphics[width=3in]{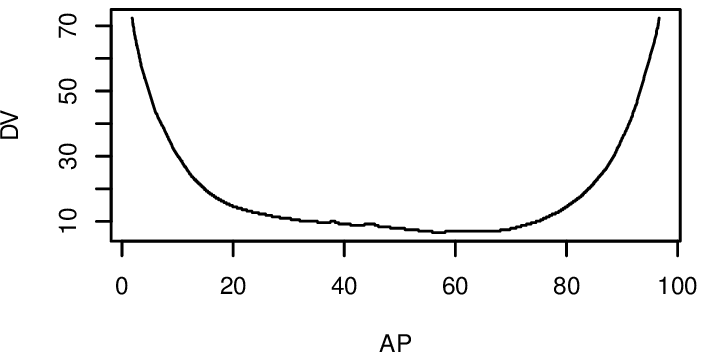}\\
\textbf{B}\includegraphics[width=3in, height=2in]{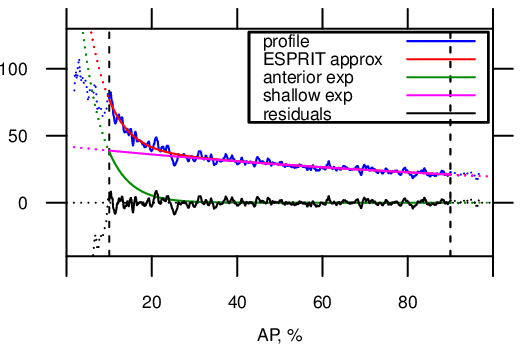}\\
\textbf{C}\includegraphics[width=3in, height=2in]{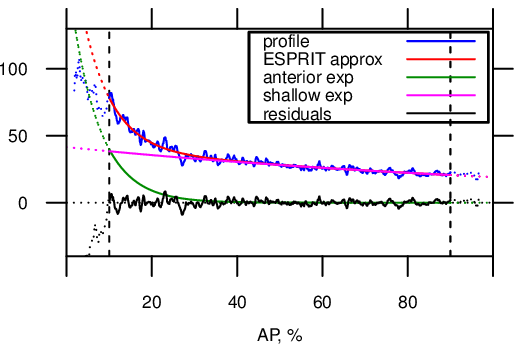}
\caption{1D profiles: curvilinear coordinates vs. AP projections. \textbf{A}. original AP, DV coordinates of the sampled nuclei. \textbf{B}. 1D intensity profile and ESPRIT analysis on an AP projection (DV coordinate not used). \textbf{C}. 1D intensity profile and ESPRIT analysis using curvilinear coordinates.}
\label{fig:curvilinear}
\end{center}
\end{figure}

Table~\ref{tab:AP_curv_mean} shows the mean values of the model characteristics for both AP and curvilinear coordinates (MSE: mean squared data-to-model difference). MSE is smaller for AP than curvilinear for all groups except CleavageEarly.
Values of $\lambda_\textrm{anterior}^\textrm{(apical)}$ are larger with curvilinear coordinates, but this stems from the way the coordinates are constructed –-- distances between points in curvilinear coordinates near the profile edges are larger than that for direct AP coordinates. The results in
Table~\ref{tab:AP_curv_mean} indicates that conclusions found using AP coordinates are valid with respect to curvilinear coordinates, and that the two-exponential model is robust to moderate deviations in the data.

\begin{table}[!htb]
\begin{center}
\caption{Mean characteristics for AP and curvilinear coordinates}
{
\begin{tabular}{|l|p{1.2cm}|p{1cm}|p{1.2cm}|p{1cm}|r|} \hline
AP-coordinates & $\lambda_\textrm{anterior}^\textrm{(apical)}$ & MSE, apical & $\lambda_\textrm{anterior}^\textrm{(basal)}$ & MSE, basal & $C^{ab}$ \\ \hline \hline
CleavageEarly & 0.79 & 15.77 & 0.78 & 13.86 & 0.11 \\ \hline
CleavageLater & 0.84 & 25.93 & 0.86 & 29.32 & 0.35 \\ \hline
nc10--12 & 0.88 & 19.17 & 0.86 & 14.35 & -0.21 \\ \hline
nc13 & 0.88 & 17.94 & 0.85 & 5.63 & -0.21 \\ \hline
nc14 early1 & 0.88 & 17.52 & 0.83 & 4.56 & 0.40 \\ \hline
nc14 early2 & 0.86 & 17.95 & 0.85 & 3.81 & 0.99 \\ \hline
nc14 late & 0.87 & 10.40 & 0.86 & 1.93 & 1.33 \\ \hline
Curvilinear coordinates & $\lambda_\textrm{anterior}^\textrm{(apical)}$ & MSE, apical & $\lambda_\textrm{anterior}^\textrm{(basal)}$ & MSE, basal & $C^{ab}$ \\ \hline \hline
CleavageEarly & 0.84 & 12.63 & 0.81 & 13.73 & 0.01 \\ \hline
CleavageLater & 0.85 & 26.83 & 0.88 & 30.89 & 0.24 \\ \hline
nc10--12 & 0.91 & 20.63 & 0.90 & 16.48 & -0.14 \\ \hline
nc13 & 0.91 & 20.53 & 0.89 & 6.73 & -0.23 \\ \hline
nc14 early1 & 0.91 & 19.11 & 0.87 & 4.99 & 0.49 \\ \hline
nc14 early2 & 0.90 & 20.10 & 0.89 & 4.10 & 1.12 \\ \hline
nc14 late & 0.90 & 11.23 & 0.92 & 2.13 & 1.59 \\ \hline
\end{tabular}
}
\label{tab:AP_curv_mean}
\end{center}
\end{table}

\end{document}